# STRUCTURAL AND MORPHOLOGICAL INVESTIGATION OF LANGMUIR-BLODGETT SWCNT/BEHENIC ACID MULTILAYERS

T. Di Luccio*, F. Antolini, P. Aversa, G. Scalia and L. Tapfer

ENEA, UTS-MAT, Centro Ricerche Brindisi,
Strada Statale 7 Appia km.713.7, 72100 Brindisi (Italy)

## Abstract

The goal of our work has been the incorporation of single wall carbon nanotubes (SWCNTs) in Langmuir-Blodgett organic multilayers. We deposited multilayers consisting of six cadmium behenate (CdBe) layers alternated with one layer of SWCNTs. SWCNTs and CdBe molecules were spread at the air/water interface and deposited at a fixed compression pressure for CdBe (27mN/m) and two different compression pressures for the nanotubes, 15 and 45mN/m, respectively. Low angle X-ray measurements exhibited distinct satellite peaks in all the samples demonstrating that the periodicity of the LB CdBe reference sample was conserved when SWCNTs were inserted in the structure. In agreement with the observations at optical and electronic microscopes, the samples deposited at the higher compression pressure (45mN/m) presented more densely-packed and more uniform coverage of nanotubes.

* Corresponding Author: Tel. +39 0831 507433; Fax. +39 0831 507674;

E-mail: tiziana.diluccio@brindisi.enea.it

# 1. Introduction

Langmuir-Blodgett technique is a powerful method to organise organic films as complex structures [1]. It allows the realisation of a molecular architecture of different molecules with high control and reproducibility over the number of repeated layers and their thickness. The resulting structure, in optimised process conditions, is well ordered with a preferential orientation.

The incorporation of "guests" in the Langmuir-Blodgett films such as atoms, molecular clusters and nanoparticles are of great scientific and technological interest due to the fact that the functional properties of the organic films and of the "guest materials" are combined together [2,3,4]. These may lead to potentially new properties and applications as electronic, optical and magnetic materials as well as sensor devices. On the other hand Langmuir-Blodgett technique allows a controlled assembling and incorporation with high spatial order of atoms, clusters and nanoparticles in the organic matrix.

In the present work we report on our first results of incorporating carbon nanotubes in Langmuir-Blodgett behenic acid films. Manipulation of SWCNTs is a major topic in the ongoing research for their technological applications. Our samples may be of interest due to their enhanced mechanical properties (toughness, stability, etc.) and for their potential application for example in SAW gas sensor technology [5].

# 2. Sample preparation and characterisation techniques

We deposited multilayers of cadmium behenate (CdBe) molecular layers and single wall carbon nanotubes (SWCNTs) layers on hydrophobic Si(100) substrates by using the Langmuir-Blodgett (LB) technique. The repetitive unit or period $D$ of the multilayers consisted of $m$ cadmium behenate layers alternated with one layer of SWCNTs. Two series of samples were prepared with $n = 5$ and $n = 10$ periods, respectively. Each of the two series was deposited with $m = 5$ and $m = 6$ CdBe layers. On top of the multilayered films $r$ CdBe layers were deposited as capping layer. In addition two different compression pressures ($\pi = 15$mN/m and 45mN/m) were used during the SWCNTs deposition. The final multilayer structure can be expressed as: ($Dn+rAB$), where $D = mAB/1SWCNT$ is one unit cell of the LB-SWCNT structure and AB indicates one LB period of the CdBe molecule that is about 30Å thick [6]. In Table 1 the deposition and structural parameters of the synthesized LB-SWCNT multilayers are summarized.

Behenic (or docosanoic) acid $C_{22}H_{44}O_2$ was diluted to 0.79mg/ml in chloroform while SWCNT raw soot was resuspended in chloroform at a concentration of 0.28mg/ml. The SWCNT solution was then sonicated for one hour in order to help the dispersion of the nanotubes. The dispersion was stable for several days. The two solutions were spread at the air/water interface on two separate compartments of the Langmuir trough (KSV 5000) containing an aqueous solution (pH=6.0) with $10^{-4}$mol divalent $Cd^{2+}$ ions dispersed in it as subphase. The films were deposited by vertical lifting at a surface pressure value of 27mN/m for CdBe while two different pressures, 15 and 45mN/m, were used for SWCNTs deposition. The CdBe molecules were both deposited during the dipping and the raising of the substrate for the samples containing an even number of CdBe layers ($m = 6$), while for those with an odd number of layers ($m = 5$) the last raising was through the neutral section. The SWCNT single layer was always deposited during the dipping.

The structural ordering and the interface configuration of the samples were investigated by X-ray reflectometry [7] performed with a Philips X'Pert Pro diffractometer. The information obtained from X-ray analysis were compared with the results of polarised light optical microscopy and scanning electron microscopy (SEM).

## 3.    Results

We started our investigations from a multilayer containing CdBe only (20 layers) and no carbon nanotubes (sample AB5). This sample has been used as reference sample in order to interpret the experimental evidences observed for the LB-SWCNT samples. Then we deposited another sample with 20 layers of CdBe as well but with one layer of SWCNTs between the first 10 and the second 10 CdBe layers (sample ABNT10). The pressure used for the SWCNTs deposition was 15mN/m. In Fig.1 we show the X-ray specular reflectivity in the measured angular range $0° - 7°$. In AB5 strong Bragg (00$l$) peaks are present whose distance corresponds to a spacing of 60.2Å. This value is in agreement with the unit cell length of the CdBe reported in literature [6]. The Bragg reflections are visible over a wide angular range, until around 35°. We also observe well defined finite thickness fringes (Kiessig fringes). Their number is equal to eight that corresponds to the total number of unit cells deposited [8]. The value of the total thickness evaluated from the distance between the minima is 607Å, in agreement with the expected value. By comparing the AB5 and ABNT10 reflectograms we observe that the 60Å periodicity of CdBe is preserved but the interface profile is modified. This demonstrates that the carbon nanotubes were deposited and incorporated in the LB film, but their presence influenced the interfaces, which is evidenced by the weak intensity and smeared-out Kiessig fringes.

In the present paper we limit our discussion to the multilayers with $m = 6$ layers of CdBe, because no relevant difference has been observed for the series with $m = 5$. Fig. 2 shows the X-ray specular reflectivity measurements of the multilayer samples of Table 1. ABNT4 and ABNT8 were deposited at the same surface pressure $\pi = 15$mN/m, but with a different number of repetitions, $n = 10$ and $n = 5$ respectively. Fig. 3a and Fig. 3b show the optical microscope images relative to these two samples. For both samples the Bragg peak positions are substantially the same with respect to the peak positions measured for the CdBe LB film (Fig. 2). There are also different features that can be related to their morphology shown in Fig. 3a and Fig. 3b. A homogeneous distribution of small dark regions and smooth zones characterises the surface of ABNT8 (Fig. 3a). The pressure of 15mN/m used for the compression of the nanotubes during their deposition is not sufficient to get a complete coverage of the CdBe surfaces. Consequently, at each period a certain number of such regions with lack of nanotubes is present. This explains why the X-ray waves are reflected with high coherence from the CdBe layers and the thickness fringes are so clearly pronounced in the sample ABNT8. The number of the fringes is related to the overall number of CdBe layers deposited. The morphology of ABNT4 is very different (Fig. 3b). The increase of repetitions of the multilayer period contributes to get higher density of SWCNTs. This leads to the disappearing of the thickness fringes in the corresponding X-ray diffractogram of Fig. 2. Here, it is worth noting that low-angle reciprocal space mapping revealed, for all the samples, a pronounced diffuse scattering close to the LB satellite peaks that gives rise to quasi-isointensity bands along $Q_x$, i.e. the momentum transfer component parallel to the surface. This clearly indicates that the LB interface configuration is highly conformal (strong correlation of the interface roughness) [9].

The influence of $n$ on the structural properties is less crucial when higher compression pressure $\pi$ is used. A better coverage by the nanotubes is found as shown by the reflectivity measurements of the two samples deposited at $\pi$ = 45mN/m (ABNT5 and ABNT9 in Fig. 2). The occurrence of changes in the interface profile of CdBe when more and more nanotubes were inserted is evident from the appearing of additional reflections located between the main satellite Bragg peaks. Moreover only very weak Kiessig fringes are detected in ABNT9 ($n$ = 5), while they are completely absent in sample ABNT5 ($n$ = 10). The pressure $\pi$ of 45mN/m is high enough to compact the molecules containing the nanotubes even when the number of periods is less than 10. This finding is confirmed by observations at the optical microscope that show morphological features similar to ABNT4 in Fig. 3b. Scanning electron microscopy was used to get a more detailed view of the features observed by the optical microscope. On the surface of ABNT4 and ABNT9 it was possible to distinguish on a scale of 20μm a dense network of nanotubes with zones where an exfoliation of the network occurred and upper and lower layers of nanotubes are visible. The nanotubes in the upper layers appear like a skein, where long bundles are densely interwoven. In ABNT9, deposited at the pressure of $\pi$ = 45mN/m with respect to $\pi$ = 15mN/m of sample ABNT4, some bundles are so much compressed that they come out of the film as well evidenced in the SEM image of Fig. 4. From our experiments we have the indication that the carbon nanotubes we incorporated in the multilayers are very likely to be too long. In fact, relatively long carbon nanotubes favour their tendency to texture densely in sheets within the CdBe layers but with a degree of coverage that is insufficient to achieve an optimized layering.

## 4. Conclusions

In this paper we reported our first attempt of incorporating carbon nanotubes within highly ordered cadmium behenate layers deposited by LB technique. Two series of multilayers were prepared in which we inserted one layer of SWCNTs every six layers of CdBe. X-ray reflectivity and microscopy observations evidenced a dependence of the layering properties on the surface pressure by which the nanotubes were deposited and on the number of repetitions of the multilayer period. Our experimental data clearly demonstrate that carbon nanotubes can be incorporated in LB films without destroying the high LB multilayer periodicity. These first results are promising for the fabrication of highly ordered carbon nanotubes–LB films composites that would exhibit new structural, mechanical and physical properties. Our future work will be focused on a better control of the preparation process with the goal to achieve a much higher ordering of the embedded carbon nanotubes within the organic film matrix.

We thank Dr. A. Vecchione of Dipartimento di Scienze Fisiche of the University of Salerno for the use of the optical microscope and Dr. P. Morales of ENEA C.R. Casaccia (Rome) for the SEM analyses.

**List of Figure Captions**

Fig. 1. Experimental X-ray specular reflectivity patterns of samples AB5 (20AB) and ABNT10 (10AB/1SWCNT/10AB) .

Fig. 2. Experimental X-ray specular reflectivity patterns of the multilayers listed in Table 1. The curves have been shifted along the vertical axis for clarity.

Fig. 3. Polarised light optical microscope images of the samples ABNT8 (a) and ABNT4 (b) whose X-ray reflectivity curves are shown in Fig. 2.

Fig. 4. SEM micrograph of the surface of ABNT9. At this magnification (image field 5μm x 5μm) the surface texture caused by the presence of the carbon nanotubes is clearly visible. Some exfoliations from the surface are well evidenced.

**Table 1**

List of the multilayer samples together with the most relevant parameters. The compression pressure $\pi$ is referred to the nanotubes.

| Sample | Surface Pressure $\pi$ (mN/m) | Unit Cell Period $D$ | Number of Periods $n$ | Number of Molecular Layers in Cap Layer $r$ |
|--------|-------------------------------|----------------------|-----------------------|---------------------------------------------|
| ABNT4  | 15 | 6AB / 1SWCNT | 10 | 6AB |
| ABNT5  | 45 | 6AB / 1SWCNT | 10 | 6AB |
| ABNT8  | 15 | 6AB / 1SWCNT | 5  | 6AB |
| ABNT9  | 45 | 6AB / 1SWCNT | 5  | 4AB |

Fig. 1

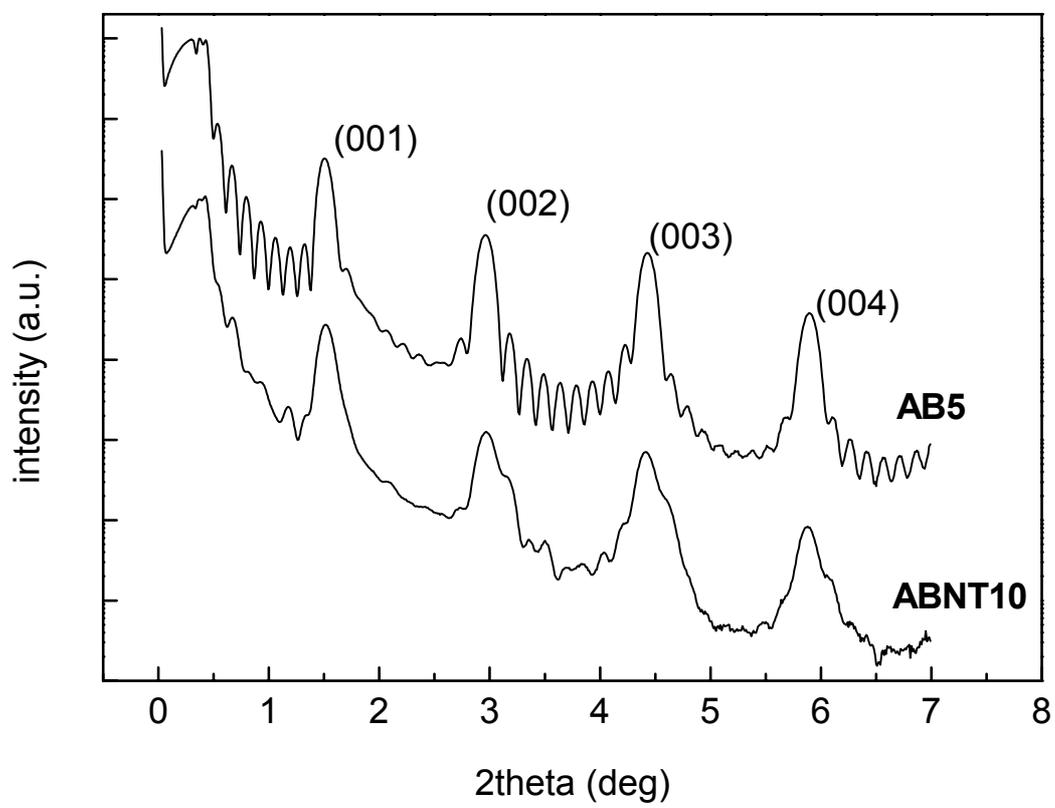

Fig. 2

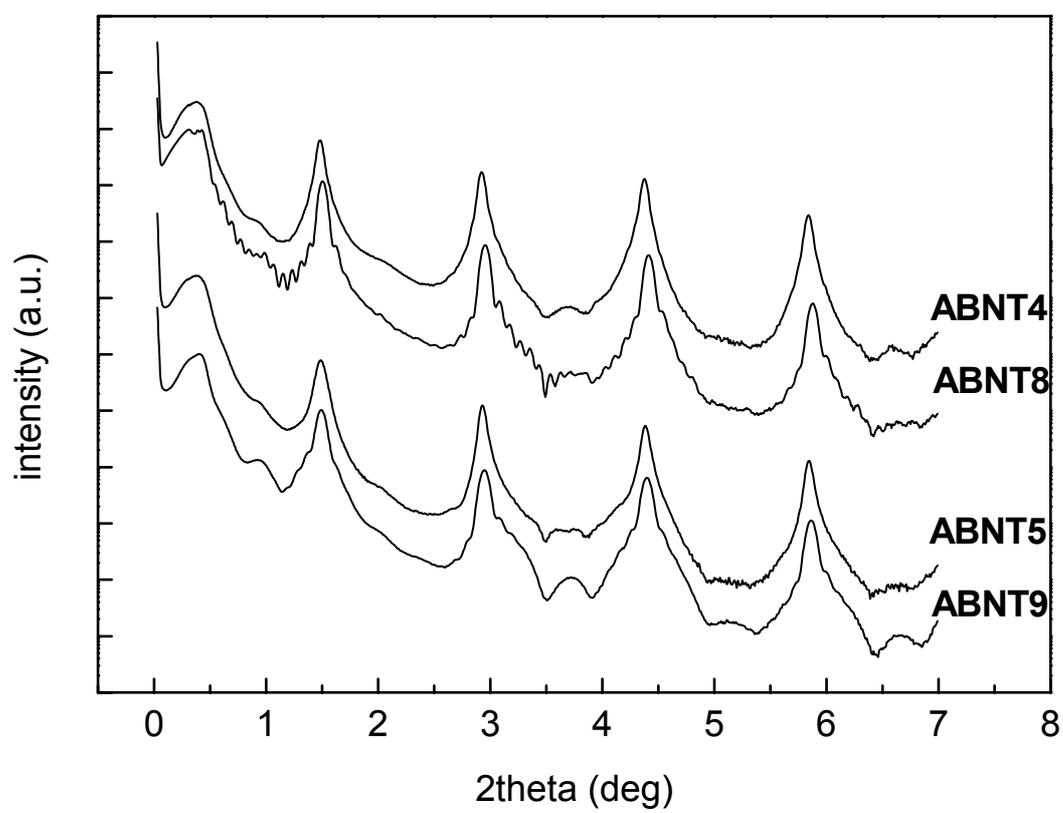

Fig. 3a                    Fig. 3b

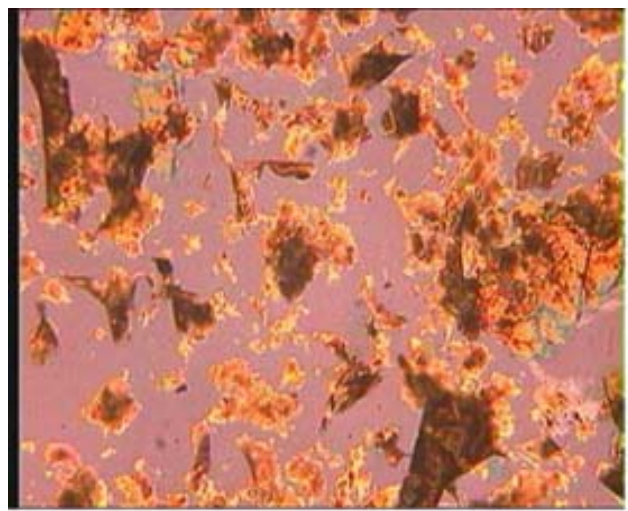 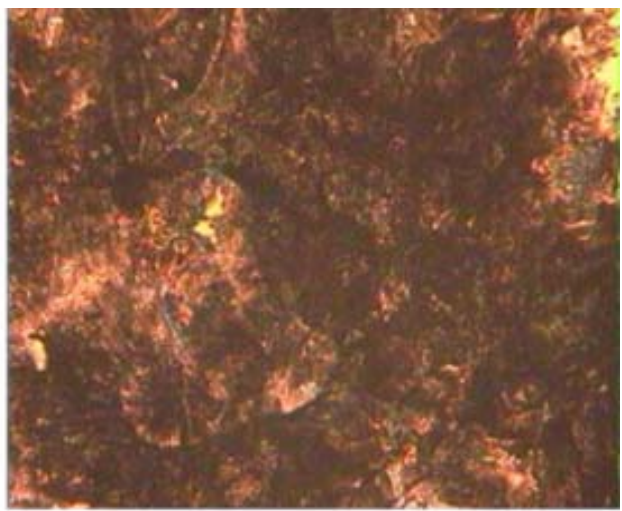

ABNT8        184μm        ABNT4        184μm

Fig. 4

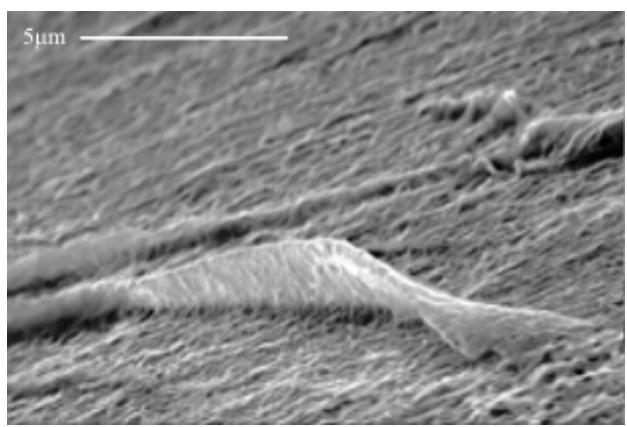